Cite as:

IEEE
M. Jafarzadeh and Y. Tadesse, "End-to-End Learning of Speech 2D Feature-Trajectory for Prosthetic Hands", in 2020 Second International Conference on Transdisciplinary AI (TransAI), 2020, pp. 25-33.

ACM
Jafarzadeh, M. and Tadesse, Y., 2020. End-to-End Learning of Speech 2D Feature-Trajectory for Prosthetic Hands. In 2020 Second International Conference on Transdisciplinary AI (TransAI). IEEE, pp. 25-33.

MLA
Jafarzadeh, Mohsen, and Yonas Tadesse. "End-to-End Learning of Speech 2D Feature-Trajectory for Prosthetic Hands". IEEE, 2020 Second International Conference on Transdisciplinary AI (TransAI), 2020, pp. 25-33, Accessed 21 Sept 2020.

APA
Jafarzadeh, M., & Tadesse, Y. (2020). End-to-End Learning of Speech 2D Feature-Trajectory for Prosthetic Hands. In 2020 Second International Conference on Transdisciplinary AI (TransAI) (pp. 25-33). IEEE.

Harvard
Jafarzadeh, M. and Tadesse, Y. (2020). End-to-End Learning of Speech 2D Feature-Trajectory for Prosthetic Hands. In: 2020 Second International Conference on Transdisciplinary AI (TransAI). IEEE, pp.25-33.

Vancouver
Jafarzadeh M, Tadesse Y. End-to-End Learning of Speech 2D Feature-Trajectory for Prosthetic Hands. 2020 Second International Conference on Transdisciplinary AI (TransAI). IEEE; 2020. p. 25-33.

Chicago
Jafarzadeh, Mohsen, and Yonas Tadesse. 2020. "End-to-End Learning of Speech 2D Feature-Trajectory for Prosthetic Hands". In 2020 Second International Conference on Transdisciplinary AI (TransAI), 25-33. IEEE.

BibTeX
@inbook{jafarzadeh_tadesse_2020,[break] title={End-to-End Learning of Speech 2D Feature-Trajectory for Prosthetic Hands},[break] booktitle={2020 Second International Conference on Transdisciplinary AI (TransAI)},[break] publisher={IEEE},[break] author={Jafarzadeh, Mohsen and Tadesse, Yonas},[break] year={2020},[break] pages={25-33}

# End-to-End Learning of Speech 2D Feature-Trajectory for Prosthetic Hands

Mohsen Jafarzadeh
Department of Electrical and Computer Engineering
The University of Texas at Dallas
Richardson, TX, USA
Mohsen.Jafarzadeh@utdallas.edu

Yonas Tadesse
Department of Mechanical Engineering
The University of Texas at Dallas
Richardson, TX, USA
Yonas.Tadesse@utdallas.edu

*Abstract*—Speech is one of the most common forms of communication in humans. Speech commands are essential parts of multimodal controlling of prosthetic hands. In the past decades, researchers used automatic speech recognition systems for controlling prosthetic hands by using speech commands. Automatic speech recognition systems learn how to map human speech to text. Then, they used natural language processing or a look-up table to map the estimated text to a trajectory. However, the performance of conventional speech-controlled prosthetic hands is still unsatisfactory. Recent advancements in general-purpose graphics processing units (GPGPUs) enable intelligent devices to run deep neural networks in real-time. Thus, architectures of intelligent systems have rapidly transformed from the paradigm of composite subsystems optimization to the paradigm of end-to-end optimization. In this paper, we propose an end-to-end convolutional neural network (CNN) that maps speech 2D features directly to trajectories for prosthetic hands. The proposed convolutional neural network is lightweight, and thus it runs in real-time in an embedded GPGPU. The proposed method can use any type of speech 2D feature that has local correlations in each dimension such as spectrogram, MFCC, or PNCC. We omit the speech to text step in controlling the prosthetic hand in this paper. The network is written in Python with Keras library that has a TensorFlow backend. We optimized the CNN for NVIDIA Jetson TX2 developer kit. Our experiment on this CNN demonstrates a root-mean-square error of 0.119 and 20ms running time to produce trajectory outputs corresponding to the voice input data. To achieve a lower error in real-time, we can optimize a similar CNN for a more powerful embedded GPGPU such as NVIDIA AGX Xavier.

## I. Introduction

The human hands are essential organs to perform daily activities ranging from hand gestures to object manipulation. The loss of a hand drastically damages humans both physically and mentally. Researchers estimated that there are approximately 94,000 upper limb amputees in Europe [1] and 41,000 upper limb amputees in the United States [2]. The World Health Organization estimates that there are about 40 million amputees in the world [3]. We expect these numbers grow because of an increase life expectancy and a corresponding higher incidence of diabetes and vascular diseases. Therefore, the idea of developing a human-like artificial hand has a long history and should continue. Developing a prosthetic hand requires great effort and multidisciplinary knowledge including but not limited to artificial intelligence, cognitive science, control, biology, neuroscience, mechanisms, sensors, and actuators. Control of prosthetic hands is essential in developing these devices. By taking advantage of the latest technological advances, engineers develop more dexterous, realistic, and novel hands. However, there are still huge gap between performance of artificial hands and human-like hands. Indeed, the development and control of a prosthetic hand is still an open problem in robotics and bioengineering. We refer readers to review papers for extensive discussion on prosthetic hands [1-8]. Here, in this paper, we discuss only the control aspects of prosthetic hands by speech 2D feature.

Electric powered prosthetic hands can receive commands from users by one or few ways such as push-button, joystick, keyboard, text, electroencephalography (EEG) [9], Electroneurography (ENG) [10-12], electromyography (EMG) [13-20], vision [21-23], and speech [24-28]. Among these ways, electromyography is the most convenient for amputees. However, electromyography cannot be used for all amputees because some of them lost their neurons. Communicating and expressing ideas via speech are easy for human. Thus, controlling prosthetic hand by speech is the most popular idea after electromyography [28]. Here, we only focus on controlling prosthetic hands via speech for the same reason. Fig. 1 shows the functional block diagram of the system discussed in this paper.

Researchers have used automatic speech recognition systems to control robots and other smart devices by speech command. Basically, automatic speech recognition systems convert speech to text. Developers can use a simple look-up table or natural language process technique to map the output text to command or trajectory. Research in automatic speech recognition systems has yielded several open-source libraries, including CMU Sphinx [29], Kaldi [30], ESPnet [31], Openseq2Seq [32], Eesen [33], and WAV2LETTER++ [34]. Zinchenko et al. [25] commanded a surgical robot by speech using CMP Sphinx. Zhang et al. [36] controlled a smart home via speech using a combination of Mel-frequency cepstral coefficients (MFCC) feature extractor, hidden Markov model classifier, and Viterbi. In another work, researchers developed a voice command module for smart home automation using long short-term memory [37]. Novoa et al. designed an automatic speech recognition to control the PR2 mobile manipulation robot using a hidden Markov model and a deep neural network [38]. Pulikottil et al. used CMU Pocketsphinx to control an assistive robotic arm [39]. Poirier et al. developed a voice control interface prototype for assistive robots using the MFCC feature extractor and Gaussian mixture model (GMM) [40]. Bingol and Aydogmus used deep neural networks to control industrial arm by speech [41]. Gau et al. controlled a hand exoskeleton by speech using a combination of discrete wavelet transforms and hidden Markov models [25]. Ismail et al. designed a speech-based controller for a robotic hand using 13 selected features that comprise eight features in the frequency domain and five features in the time domain along with a multilayer perceptron (MLP). They performed training and testing in an identical room to reduce the influence of external noise [24]. Vijayaragavan et al. used Google speech-to-text to control prosthetic hand by android devices [26]. Alkhafaf et al. used a combination of the Mel-frequency cepstral and Gaussian mixture models to control prosthetic hand [27].

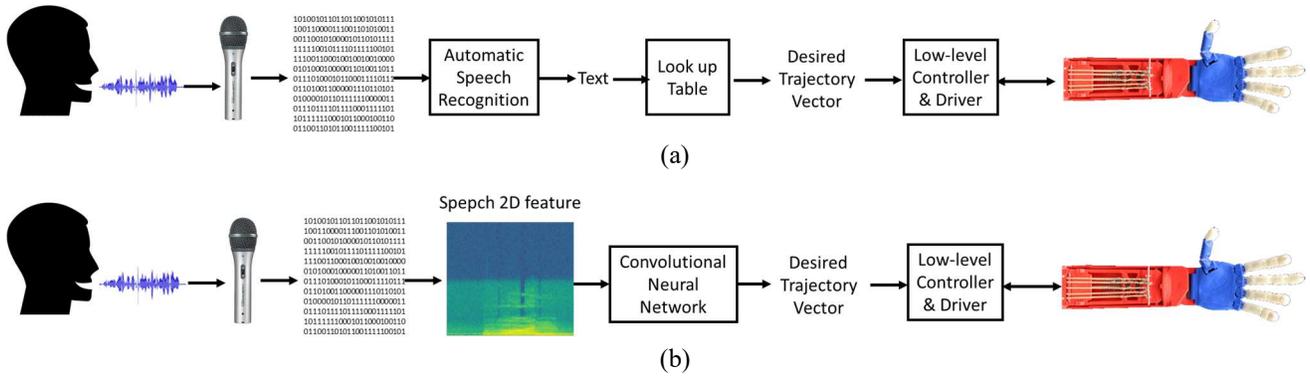

Fig. 1. Block diagram of controlling prosthetic hand via speech using (a) automatic speech recognition systems and (b) proposed method.

Advancement in general-purpose graphics processing units (GPGPUs) [42-44] enables mobile devices to run a large set of computations in parallel (real-time) [45]. Moreover, a huge amount of labeled speech data has been released in the past few years. Thus, researchers can train very deep neural networks and use them in real-time scenarios, by using these recent parallel computing processors and big data. As a result, the state-of-the-art optimization methods quickly changed from subsystem optimization to end-to-end optimization [46, 47]. Image segmentation [48], object detection [49], speech recognition [50], natural language processing [51], playing Atari game [52], object manipulation [53], autonomous driving [54], lung disease detection [55], and financial assert trading [56] are some of the examples of the optimization paradigm shift.

In this paper, we explain an end-to-end learning of speech 2D feature to trajectory for controlling prosthetic hands by using speech input (Fig. 1). Here, a trajectory is defined as a time-invariant vector that its elements correspond to positions of each mechanical degree of freedom of the hand. For example, for a hand with 5 fingers and 5 actuator, the trajectory is a vector with 5 elements with values ranging from 0 (full relaxation) to 1 (full contraction). We investigate CNNs that directly maps the speech 2D feature to trajectory and run in real-time on an embedded GPU. We omit the speech to text steps in controlling the prosthetic hand. The main advantage of the proposed method is that the CNN learns useful features and is optimized for trajectory instead of text. Another advantage of this approach is that the trajectory is surjective function, not only from speech 2D features but also from the speech dictionary. The speech dictionary is defined as the set of all words that the network is trained with such as "open", "close", "one", "off", etc.

The paper is organized as follows. First, an end-to-end learning of speech 2D feature-trajectory is explained for prosthetic hands with convolutional neural networks (CNN). Then, computational results of the proposed method on an embedded GPGPU are presented. Discussions about the proposed method are stated in the next section. The last section is a conclusion on the proposed method for controlling prosthetic hands with embedded GPGPU by speech command.

## II. METHOD

In this section, we describe a new method to control prosthetic hands with speech, but without using an automatic speech recognition. Fig. 1 shows the overall system. First, a digital microphone (microphone with an embedded analog to digital converter (ADC)) reads the speech signal (sampled at 16 kHz or higher) as an integer (16-bit or higher). If the ADC output is a float-point or a fixed-point number, we would map it linearly to an integer number. Second, a GPGPU kit computes the speech 2D features from the stream of the sampled speech signal. The proposed method can use any speech 2D features that have a local correlation in each dimension. Spectrograms, the logarithm of spectrograms, Mel-frequency cepstral coefficients (MFCC), and power-normalized cepstral coefficients (PNCC) are examples of speech 2D features. In this method, we use a lightweight CNN that maps a speech 2D feature to the corresponding trajectory (vector). The trajectory is defined as a time-invariant vector that its elements correspond to positions of each mechanical degree of freedom. Finally, the trajectory vector is sent to the low-level controller and driver of a prosthetic hand.

Here, we assume that prosthetic hands have 5 degrees of freedom corresponding to each finger. Therefore, the network output (vector) has 5 elements. This method is generalizable to

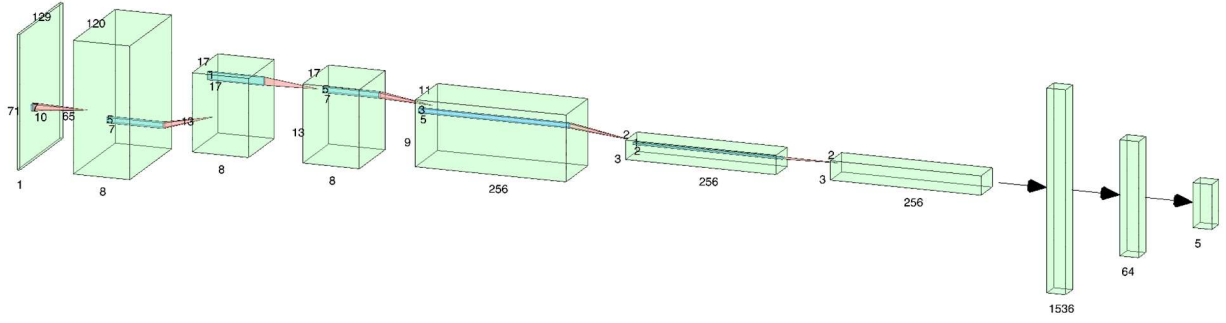

Fig. 2. Block diagram of the proposed convolutional neural network that optimized for NVIDIA Jetson TX2 developer kit (details stated in table 1).

TABLE I. SUMMARY OF THE PROPOSED CONVOLUTIONAL NEURAL NETWORK THAT OPTIMIZED FOR NVIDIA JETSON TX2 DEVELOPER KIT

| Layer | Type | # filters | Filter size | Stride | Activation | Output shape | # Parameters |
|---|---|---|---|---|---|---|---|
| 0 | Input (Log of spectrogram) | - | - | - | - | 129 x 71 x 1 | 0 |
| 1 | Convolution 2D | 8 | 10 x 7 | 1 | ReLU | 120 x 65 x 8 | 568 |
| 2 | Max Pooling 2D | - | 7 x 5 | 7 x 5 | Max | 17 x 13 x 8 | 0 |
| 3 | Batch normalization | - | - | - | - | 17 x 13 x 8 | 32 |
| 4 | Convolution 2D | 256 | 7 x 5 | 1 | ReLU | 11 x 9 x 256 | 71936 |
| 5 | Max Pooling 2D | - | 5 x 3 | 5 x 3 | Max | 2 x 3 x 256 | 0 |
| 6 | Batch normalization | - | - | - | - | 2 x 3 x 256 | 1024 |
| 7 | Flatten | - | - | - | - | 1536 | 0 |
| 8 | Dense | - | - | - | ReLU | 64 | 98368 |
| 9 | Drop out | - | - | - | - | 64 | 0 |
| 10 | Dense | - | - | - | ReLU | 5 | 325 |

prosthetic hands with a higher or lesser degree of freedom. The trajectory is defined as

$$\vec{T} = \begin{bmatrix} r_{thumb} \\ r_{index} \\ r_{middle} \\ r_{ring} \\ r_{pinky} \end{bmatrix} \in [0,1]^5 \subset \mathbb{R}^5 \quad (1)$$

Each element of the trajectory can be a real number ranging from 0 to 1. The value zero means completely open (full relaxation), and one means completely close (full contraction). Partially open finger positions have intermediate values. The speech dictionary is defined as the set of all words that the network is trained with such as "zero", "one", "two", "off", etc. In this approach, the output size is fixed regardless of the number of speech dictionary. For example, the word "two" corresponds to [1 0 0 1 1] and the word "ok" can be mapped to [0.2 0.7 0 0 0]. The main advantage of this approach is that the trajectory is surjective map not only from speech 2D features but also from the speech dictionary. The loss function is the difference between the desired and inferred trajectory.

$$\vec{\ell} = \vec{T}^{desired} - \vec{T}^{inferred} \quad (2)$$

In optimization literature, many error functions were proposed such as mean absolute value (Lasso), maximum absolute value, logarithm of cosine hyperbolic, Tuckey, Huber, etc. In this method, the error of the network for a 5 finger hand is defined as the mean squared (MSE) of loss functions.

$$MSE = \frac{1}{5N} \sum \vec{\ell}^T \vec{\ell} \quad (3)$$

$$RMSE = \sqrt{\frac{1}{5N} \sum \vec{\ell}^T \vec{\ell}} \quad (4)$$

Where T is transpose operation, and $N$ is total number of training/validation data. The mean squared function gives more weight to larger loss than the loss near the origin. Therefore, the training process would result in a smaller maximum loss in prosthetic fingers than the methods with linear function near the zero, such as mean absolute error. Minimizing the maximum absolute error produces better results than minimizing the mean square error because in many practical applications maximum error should be bounded. If we minimize the mean square error, some speech commands can have a large error to reduce small errors of the speech commands. However, currently, we could not find a way to train the network with this function because we trained the network with gradient descent optimization family (Adam) and we cannot find a stable and robust numerical or analytical method to find gradient of maximum absolute error.

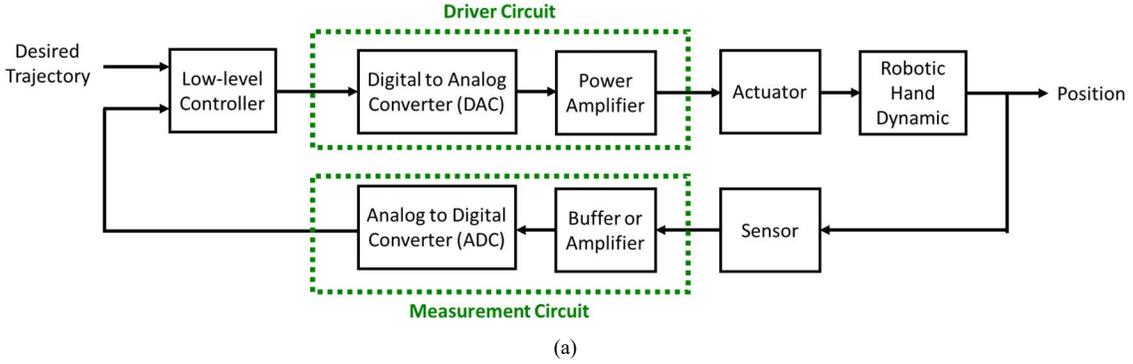

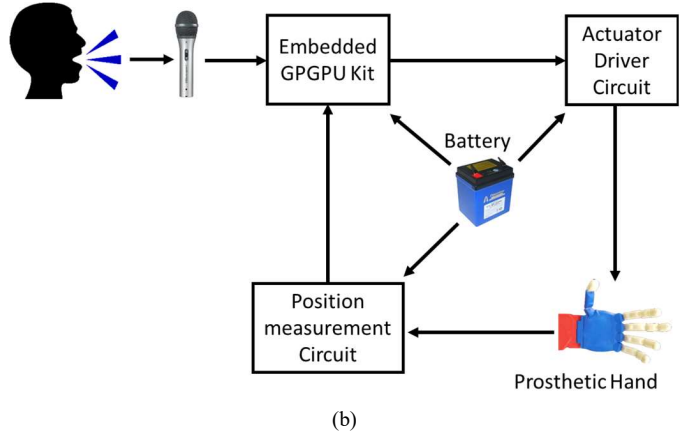

Fig. 3. Diagram of a general low-level control system (a) function block diagram and (b) hardware diagram.

To map the speech 2D feature to the corresponding trajectory, we investigated convolutional neural networks because it is easier to train CNN than dense neural networks (DNN) and fully connected recurrent neural networks. Fig. 2 shows the tensor diagram of the proposed CNN. Here, we use 129 x 71 speech 2D feature (matrix) as the input. The first layer is a 2D convolutional layer with 8 filters (10 x 7), one stride, and activation function type of rectified linear units (ReLU).

$$\text{ReLU}(x) = \begin{cases} 0, x \leq 0 \\ x, x > 0 \end{cases} \quad (5)$$

where x is a real number. The second layer is 2D max-pooling with the size of 7 x 5 and stride of 7 x 5. The third layer is batch normalization. The fourth layer is a 2D convolutional layer with 256 filters (7 x 5), stride of 1, and ReLU activation function. The fifth layer is 2D max-pooling with the size of 5 x 3 and the stride of 5 x 3. The sixth layer is batch normalization. The data is flattened to a 1536 elements vector in the seventh layer. The eighth layer is a dense (fully connected) layer with ReLU activation function and output size of 64. To improve the training, the ninth layer is the drop out layer. Finally, the last layer is a dense (fully connected) layer with the ReLU activation function and output size of 5. The number 5 corresponds to the degree of freedom of the prosthetic hand, i.e., the five fingers. The proposed network has 171,725 trainable parameters. A summary of the proposed convolutional neural network is stated in Table I. To determine the optimal size of filters and to keep trade-off between speed and accuracy, we start from the values in our past work [28] and manually search the better parameter using heuristic search (trial and error). In the future, we will use an automatic grid search to find the best feature size.

After converting the 2D speech feature to a trajectory, a low-level controller will be used to guide the prosthetic hand to the desired trajectory. The low-level controller reads the trajectory and current positions of the fingers and applies control law (voltage) by the driver circuit (Fig. 3). The low-level controllers can be a PI/PID controller, Fuzzy controller, model-predictive controller, etc.

III. EXPERIMENTAL SETUP AND SPEECH DATA SET

To test the proposed method, we created an experimental setup (Fig. 3b). For capturing the speech signals, we used a USB digital microphone (Audio-Technica ATR2100-USB). We designed two PCB. First PCB is driver board that consists of a 16-bit digital to analog converter (DAC) and power amplifiers. The second PCB is measurement board that consist five 16-bit Sigma-Delta analog to digital converter (ADC) and five buffers (Op Amp with gain 1). The custom-designed PCBs communicate with a GPGPU kit via an I2C bus. Detail of these PCBs published on our previous paper [13, 28].

Currently, there are few embedded GPGPU in the market. Table II states two models along with their specifications. Here, we selected NVIDIA Jetson TX2 developer kit for the sake of cost. The proposed CNN has been optimized for NVIDIA Jetson TX2. Indeed, users can run the proposed CNN without any change in the NVIDIA AGX Xavier developer kit. The kit runs

TABLE II. EMBEDED GPGPU DEVELOPER KITS COMPARISON

| Company | NVIDIA | NVIDIA |
|---|---|---|
| Model | Jetson TX2 | AGX Xavier |
| GPU | Pascal 256 core | Volta 512 core |
| CPU | 4 core Cortex-A57 + 2 core Denver | 8 core Carmel |
| RAM | 8 GB | 16 GB |
| Storage | 32 GB | 32 GB |
| TFLOPS | 0.559 | 1.3 |
| GPIO | 8 | 4 |
| USB | 1 x USB 3.0 +1 x USB 2.0 | 2 x USB C [3.1] |
| UART | 1 | 1 |
| I2C | 4 | 2 |
| SPI | 1 with 2 CS | 1 with 2 CS |
| CAN | 1 | 1 |
| I2S | 2 | 1 |
| Size (mm) | 170 x 170 x 51 | 105 x 105 x 85 |
| Weight | 1.5 Kg | 630g |
| Price ($) | 400 | 700 |

TABLE III. NUMBER OF UTTERANCES IN GOOGLE SPEECH COMMAND DATA SET

| Words | # Training Data | # Validation Data 1 | # Validation Data 2 | Total |
|---|---|---|---|---|
| Zero | 3250 | 384 | 418 | 4052 |
| One | 3140 | 351 | 399 | 3890 |
| Two | 3111 | 345 | 424 | 3880 |
| Three | 998 | 356 | 405 | 1759 |
| Four | 2955 | 373 | 400 | 3728 |
| Five | 3240 | 367 | 445 | 4052 |
| On | 3086 | 363 | 396 | 3845 |
| Off | 2970 | 373 | 402 | 3745 |
| Other words | 62093 | 7069 | 7716 | 76878 |
| Total | 84843 | 9981 | 11005 | 105829 |

Linux Ubuntu 18.04 operating system. In the current experimental setup, we used Python programming language with NumPy, SciPy, and Keras (with the backend of TensorFlow) libraries. If a higher speed is required in a specific application, we can rewrite the programming codes in C and C++ with TensorRT (deep learning inference optimizer for NVIDIA GPGPU kits).

In this paper, we used the logarithm of spectrogram for speech 2D feature. First, we calculated the spectrogram using the SciPy library. To avoid singularity in the next steps, a negligible number ($10^{-10}$) has been added to the spectrogram. Then, we used the NumPy library to calculate the logarithm from the previous step to compute a speech 2D feature. We postponed the study of other types of speech 2D features to future works.

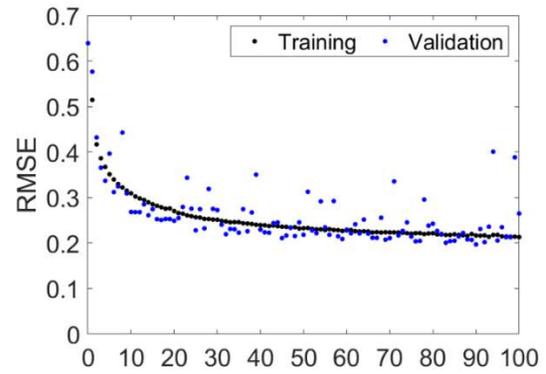

(a)

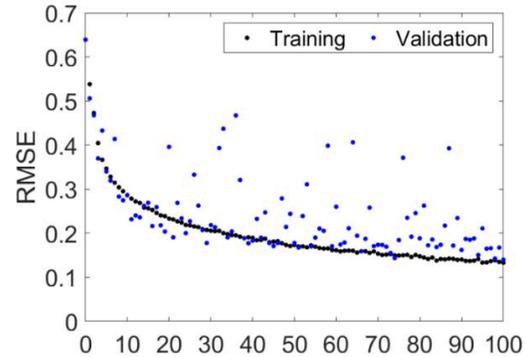

(b)

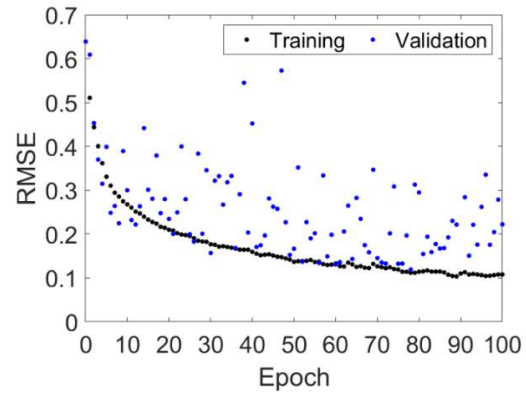

(c)

Fig. 4. Root mean square error (RMSE) of proposed end-to-end convolutional neural networks (Table 5.10) with (a) 32, (b) 128, and (c) 256 filters in second convolutional layer using Google speech command data set. The validation root-mean-square errors are 0.153, 0.140, and 0.119 in order.

Data set is an essential part of the supervised learning of models. Especially, deep convolutional neural networks require a huge amount of data for training and validation. In these experiments, an open-source Google speech command data set [57] is used. Google released this data set under the Creative Commons BY 4.0 license. This data set has 35 words (total 105,829 utterances from several people). Google saved each utterance as a WAV format file. Each utterance is one-second or shorter. The speech files encoded in linear 16-bit single-channel PCM values with 16 kHz sample rate. Google also added several minutes long audio files with various kinds of background noise.

They captured these background noises directly from noisy environments. To achieve a more robust network to environmental noise, the speech data can be augmented with the stored environment noise and use the augmented data for training. Here, we trained the networks with 8 words from 35 words, which are "zero", "one", "two", "three", "four", and "five", "on" and "off". The number of data for training is $N = 22,750$ for training words "zero" to five, "on", and "off" (Table III). The rest of the words on the data set mapped to zero vector (fully relaxed) for safety.

## IV. RESULTS

We minimized the size of CNN to let it run in real-time on NVIDIA Jetson TX2 model. We trained the networks on NVIDIA Jetson TX2, i.e., we did not use an external GPGPU. Fig. 4 demonstrates the trend of training error along with the validation error (validation data 1 in Table III) of three different trials on the Google speech command data set. Each epoch takes about 90 s. The CNN has the same structure as Table I except the number of filters in the second convolutional layer (the fourth layer in Table I), which are 64, 128, and 256. The weights (parameters) of networks are initialized by random numbers generated by Glorot uniform (also called Xavier uniform) function distributions. For the training of each CNN, we used the Adam optimizer. The objective of optimization is the mean squared (MSE) of loss functions shown in (3). This objective function also minimizes the RMSE because there is a bijective mapping from MSE to RMSE. This optimization is limited by RAM of the NVIDIA Jetson TX2. In the CNN architecture, for the number of filters corresponding to 64, 128, and 256, the root-mean-square error is 0.153, 0.140, and 0.119, respectively. This RMSE is obtained by getting the second root from the minimum MSE of validation. The minimum MSE of training is always the last point while the minimum MSE of validation may be located in the middle of training. For example, the RMSE of 0.119 belongs to the 78'th point of Fig. 4c.

The smaller the error, the better the accuracy of predicting a particular gesture. To get more accuracy in real-time, we must use a more powerful embedded GPGPU, i.e., NVIDIA AGX Xavier such that we can increase the number of filters and layers. The text (words) error is not defined and not reported here because the proposed method is end-to-end learning and we did not classify the speech to text (even in the middle step). Further work can be done to compare this approach with others and considering other performance metrics. The proposed network runs in real-time, 20 to 40 ms, on the NVIDIA Jetson TX2 developer kit.

As a preliminary test, the proposed end-to-end method was tested on a prosthetic hand/ robotic hand that has 5 fingers on a tabletop test. As shown in Fig.5, a subject uttered words such as "zero", "five", and "open", the spectrogram was showed the same as the spoken words correctly and the robotic hand was actuated according to the words. For example, when the word "one" was spoken, the output of the network showed a vector of [0.979, 0, 0.983, 0.983, and 0.986] within 20.4 ms, which corresponds to the unactuated index finger and full flexion in all the rest of the fingers. This work needs more experiments on different users, number of trials and different words.

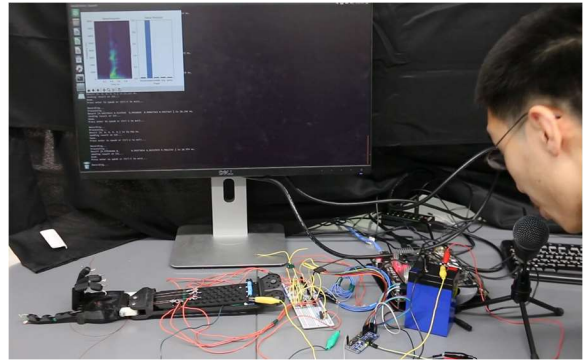

Fig. 5 Preliminary experiment on a prosthetic / robotic hand.

## V. DISCUSSION

As mentioned in previous sections, the advantage of the proposed method is that the inferred trajectory is surjective from the speech dictionary. Therefore, the user can retrain the proposed network with new data without changing the structures. In other words, the number of parameters remains the same when we add more words. However, if the number of words increases significantly, the user may need to increase the number of filters. The proposed number of filters is optimized for the 8 selected words. Our computational results show that increasing the depth of the network has little effect on overall error.

In our computation, we used the logarithm of spectrogram as speech 2D feature. When the signal to noise ratio is greater than 10, the spectrogram works well. However, the spectrogram is sensitive to noise. To have a high signal to noise ratio, we used a cardioid polar pattern microphone. Studies show that power-normalized cepstral coefficients (PNCC) is more robust than spectrogram [58]. In the future, we will compare errors resulted from the logarithm of spectrograms, MFCC, and PNCC.

There is a trade-off between speed and accuracy of CNNs. The root-mean-square error of the CNN that is optimized for NVIDIA Jetson TX2 is 0.119, with the running time of 20-40 ms. This result is the 78'th point in Fig. 4 and it is obtained by (4) and Adam optimization of the proposed network (Table I) on the Google speech command data set. We can find a CNN with a lower error if we scarify the speed. The boundary of the trade-off will be shifted by using a more powerful GPGPU such as NVIDIA AGX Xavier.

Also, there is an error trade-off between generalization and personalization. In our experiments, we trained the networks only with Google data set. Therefore, the network is generalizable. We can personalize it by fine-tuning. The operators can record their speeches and retrain the pre-trained network. By fine-tuning, the error of the networks for a specific person will be reduced. At the same time, the overall error will increase.

In this work, we trained the convolutional neural networks with words. The proposed method can be generalizable for sentences. However, the number of data for training sentences expand exponentially. Therefore, with the current amount of data, we cannot train sentence.

The generated trajectory in this method looks like a stair. i.e., it changes suddenly after running the network and remains constant up to the next running. Low-level controllers can smooth the trajectory (for example pole-placement or model reference controller). Therefore, the proposed method is useful for static hand position/gesture. This method cannot handle dynamic commands like the word "wave".

The running time on the NVIDIA Jetson TX2 developer kit is 20 to 40 ms when we used Python programming language with Keras (backend TensorFlow) library. We can reduce the running time using either WAV2LETTER++ [34] or TensorRT [42] library in C++ programming language. TensorRT decreases the accuracy of the system a little while it increases the speed. Here, we did not use TensorRT because the root-mean-square error is already high (0.119). Because the run-time is less than 50 ms, the latency is noticeable, i.e., the proposed method is functional in real-life applications.

The traditional methods use automatic speech recognition to map speech to text. They use CNN to classify audio to one-hot vectors (bag of words) or words embedding vector. Therefore, always there is a relationship between the meaning of the text (language) and prosthetics trajectory. These relationships are defined either in a lookup table or natural language processing units. However, the proposed end-to-end does not map speech to text. The proposed CNN is not a classifier. It is a regressor. As a result, there is no relationship between the words (human language as text) and the output trajectories that would be applied to the prosthetic hand. More information about the difference between classification and regression are stated in [59-62].

Our past works demonstrated that the traditional method (automatic speech recognition) could be run in 2 ms, which is a lot faster than the end-to-end method [28]. This is because the end-to-end method requires more filters (256 in comparison to 32) and consequently more computation to map speech to trajectory instead of class. Therefore, the speed (inference time) is not advantage of the end-to-end method in comparison of automatic speech recognition. The main advantage of the proposed method in comparison to the traditional method is that the network learns useful features that are directly related to trajectory instead of learning features related to the text. A good feature learning and feature representation is essential in open set recognition and open world learning [63]. In the future works, we will study the effect of feature representation of proposed end-to-end method in novelty discovery and adaptation.

The proposed convolutional neural network outputs fingers closing rates in a range of 0 to 1. Compared to motor commands, this allows transferability to different hands. Therefore, the method is not specific to the trained hand. In some cases, the size of the hand and its kinematic structure (underactuated finger coupling, thumb circumduction) do influence the adequate finger closing rates for the same gesture. This can be easily addressed by a simple linear or non-linear mapping function.

Some gesture requires binary states, for example, the word "two". These words work without fine-tuning with any hand. If an application is limited to these words, a pre-trained CNN can be downloaded and be used directly. However, in most applications, the prosthetic hands require intermediate states, for example, the word "ok". The positions of these words depend on the kinematic structure of the prosthetic hands. The proposed method can generate any value between 0 to 1 for each finger. In these applications, prosthetic hand designers can download a pre-trained CNN and freeze all the layers of CNN, except the last dense layer. Then, according to the specific kinematic structure of the prosthetic hand, the designers can retrain the last dense layer. This approach is known as transfer learning in machine learning research. In some cases, prosthetic hand designers may want to add new words that do not exist in the pre-trained model. Adding the new words depends on the quality of feature representation and the correlation of required features with learned features. Therefore, the designers can try the transfer learning approach as mentioned above, and test the accuracy of the new words. If the accuracy is not satisfactory, it means that the required features for those words did not gain during the initial training, i.e., pre-trained CNN. In this situation, the designer should collect many audios of those words from many people and retrain the network.

In this paper, we showed speech to trajectory mapping. It is functional if only the amputee commands the hand. If another person in the room says a word, the prosthetic hand reacts to it. This issue also exists in traditional methods. To solve this issue, a speaker recognition network should be added in parallel to the proposed speech to trajectory CNN. Many functional speaker recognition systems already exist, and prosthetic hand designers can add them easily [64, 65]. The speaker (amputee) recognition unit does not exist in EMG based control of prosthetic hands. As mentioned before, for some cases amputee, EMG cannot be used because that EMG cannot be accessible because of poor stump conditions or loss of neural peripheral activity. Another practical issue is that the proposed method required huge amounts of training data for adding each word. Also, this would be very difficult to obtain the necessary use cases from each specific amputee for fine-tuning. The issue also exists in EMG based control. In the future works, we will try to address the data collection issue by adding self-supervised learning [66].

## VI. CONCLUSION

In this paper, we proposed an end-to-end learning of speech 2D feature-trajectory for controlling prosthetic hands using speech with convolutional neural networks on an embedded GPGPU in real-time. First, a digital microphone captures a speech signal. Then, the speech 2D feature is computed from the speech signal stream. The voice-print is a 2D feature that has local correlations in each dimension such as spectrogram, MFCC, or PNCC. In the proposed method, we omit the speech to text steps in controlling the prosthetic hand. The proposed convolutional neural network maps the speech 2D feature to a trajectory. The trajectory is a vector with 5 elements (corresponding to each finger of prosthetic hands) with values ranging from 0 (full relaxation) to 1 (full contraction). Finally, the trajectory is sent to the low-level controller of the prosthetic hand. The main advantage of the proposed method is that the convolutional neural network learns useful features and be optimized for trajectory instead of text. Another advantage of this approach is that the trajectory is surjective not only from speech 2D feature but also from the speech dictionary.

In our experiments, we used the logarithm of the spectrogram as speech 2D features. Then, we minimized the size of convolutional neural networks such that it can be run in real-time on the NVIDIA Jetson TX2 developer kit. The proposed method implemented by Python programming language using NumPy, SciPy, and Keras with TensorFlow backend. The optimized network for NVIDIA Jetson TX2 developer kit runs in real-time (20 to 40 ms) with the root-mean-square error of 0.119. In the future, we will replace NVIDIA Jetson TX2 developer kit with NVIDIA AGX Xavier or NVIDIA Orin developer kit and optimize a similar CNN to achieve a lower error. Moreover, we will test other types of speech 2D features such as MFCC and PNCC. In addition, we will try to find a new end-to-end CNN architecture that directly maps raw speech signals to trajectories, i.e end-to-end audio-trajectory network. Experiments will be done on prosthetic hands to determine the performance and other practical issues in the future.

## ACKNOWLEDGMENT

We would like to thank Akshay Chitale, Rodrigo Avila Escobedo, Clarissa Renee Curry, and Cameron Ford for valuable contributions in developing this project. We would like to express our very great appreciation to Eric Deng, Dr. Marco Tacca, Dr. Neal Skinner, and Dr. John H. L. Hansen, Dr. Nicholas Gans, Dr. Carlos Busso-Recabarren and Dr. Mehrdad Nourani for comments and guidance that greatly improved the projects.